\documentclass[12pt,a4paper]{article}
\usepackage[utf8]{inputenc}
\usepackage[T2A]{fontenc}
\usepackage[english,russian]{babel}
\usepackage{amscd,amssymb}
\usepackage{amsfonts,amsmath,array}
\usepackage[dvips]{graphicx}
\usepackage{longtable,wrapfig}
\usepackage{lipsum}

\newcommand\summaryname{Abstract}
\newenvironment{Abstract}%
    {\small\begin{center}%
    \bfseries{\summaryname} \end{center}}

\usepackage{environ}
\usepackage{expl3}
\ExplSyntaxOn
\seq_new:N \l_riad_bib_seq % bibliography, split at \bibitem commands
\tl_new:N \l_riad_pre_tl   % part before the first \bibitem
\prg_new_conditional:Npnn \riad_str_compare:nNn #1#2#3 { TF }
  {
    \if_int_compare:w
        \pdfstrcmp { \exp_not:n {#1} } { \exp_not:n {#3} } #2 \c_zero
      \prg_return_true:
    \else:
      \prg_return_false:
    \fi:
  }
\NewEnviron{sortbibliography}[2]
  {
    \begin{#1}{#2}
      \seq_set_split:NnV \l_riad_bib_seq { \bibitem } \BODY
      \seq_pop:NN \l_riad_bib_seq \l_riad_pre_tl
      \tl_remove_all:Nn \l_riad_pre_tl { \par } % Error checking: only \par and
      \tl_remove_all:Nn \l_riad_pre_tl { ~ }    % spaces before first \bibitem.
      \tl_if_empty:NF \l_riad_pre_tl            % Otherwise, complain about the
        {                                       % tokens found before \bibitem.
          \msg_error:nnxxx { riad } { junk-before }
            { \token_to_str:N \bibitem }
            { sortbibliography }
            { \tl_to_str:N \l_riad_pre_tl }
        }
      \seq_sort:Nn \l_riad_bib_seq
        {
          \riad_str_compare:nNnTF {##1} > {##2}
            { \sort_return_swapped: }
            { \sort_return_same: }
        }
      \seq_map_inline:Nn \l_riad_bib_seq { \bibitem ##1 }
    \end{#1}
  }
\msg_new:nnn { riad } { junk-before }
  { Extra~'#3'~before~the~first~'#1'~in~environment~'#2'. }
\ExplSyntaxOff

\begin{document}

\renewcommand\refname{\centering \textit{\small{Литература}}}

\begin{center}
{\large\bf Перспективы и  ограничения нейросетевых моделей в нейронауках} \\[5mm] А.\,А.~Онучин \\
 МГУ, Москва, Россия \\
\end{center}
\begin{abstract}
Нейросетевые модели — стремительно набирающий популярность метод исследования и описания сложных мозговых процессов. Именно поэтому вопрос биологического правдоподобия и реалистичности подобных моделей является крайне актуальным. Даже с учётом постоянного прогресса в нейросетевых науках, далеко не везде удалось достичь должного уровня точности и правдоподобия в сочетании с необходимым уровнем общности описания нейробиологических механизмов. В данной работе мы обсудим существующие нейросетевые модели: локальные, аттракторные, многослойно-персептронные и глубокие, а также выделим условия в которых их биологическая правдоподобность максимальна. Вариативность условий представлена как различными нейронными моделями, механизмами пластичности и обучения, так и собственно имплементацией механизмов ингибиции и систем контроля, а также выбором сетевой архитектуры (модулярности, связности). Мы описали наиболее значимые, на наш взгляд, направления в области нейросетевого моделирования, а также потенциальные ограничения и направления развития существующих моделей.
\\
\\
\textbf{Ключевые слова:}
\textit{искусственные нейросети, моделирование сложных систем, вычислительные нейронауки}
\end{abstract}

\begin{center}
{\large\bf Perspectives and constraints on neural network models of neurobiological processes}  \\ A.\,A.~Onuchin \\
 MSU, Moscow, Russia \\
\end{center}

\begin{Abstract}
\begin{changemargin}{1cm}{1cm}
Artificial and natural neural network models are a new toolkit which could be potentially have been used for clarifying of complex brain functions. To attend this goal, such models need to be neurobiologically realistic. However, although neural networks have advanced keenly in recent decades their strict similarity in aspects of brain anatomy and physiology is imperfect. In this work we discuss different types of neural models, including localist, attractor and deep network models,  and also identify aspects under which their biological credibility can be improved. These conditions range from the choice of neuron models and of mechanisms of synaptic plasticity and learning to implementation of inhibition and control, along with network architectures (modularity, connectivity). We highlight recent advances in biologically inspired neural network models and their constraints.
\\
\\
\textbf{Key words:}
\textit{neural networks, complex systems modeling, computational neuroscience}
\end{changemargin}
\end{Abstract}

\section{Нейросетевые модели}

\subsection{Локальные нейросетевые модели}

В классических когнитивных теориях, стандартным методом построения моделей различных процессов являлся так называемый `блочный' метод. Это идея, навеянная компьютерной архитектурой фон Неймана \cite{VonNeumann1945}, укоренилась на десятилетия в репертуаре большинства исследователей в области когнитивных нейронаук. Подобные `блоки' являлись чем-то вроде модулей в нервной системе и ассоциировались со специализированными когнитивными функциями и процессами (например, звуковое восприятие), а стрелки между ними указывали на направление и порядок информационных взаимодействий. Подобные модели обычно получались на основе данных поведенческих корреляционных исследований со здоровыми испытуемыми - например, в задачах по исследованию различий восприятия речи и других звуков, это послужило основой для выделения различных модулей восприятия речи и восприятия нелингвистических акустических стимулов \cite{Liberman1967, Fodor1983}. Хотя некоторые из этих моделей были уточнены в исследованиях на людях с неврологическими нарушениями \cite{Shallice1988, Ellis2013}, большая их часть формулировалась без какой бы то ни было оглядки или ссылки на действительную нейробиологию мозга, совершенно игнорируя необходимость согласования подобного рода моделей с биологическим фундаментом \cite{Braitenberg1978, Hebb1949,  OReilly1998}.

Значимым шагом к более строгой формулировке моделей такого типа было создание так называемых \textit{локальных нейросетевых моделей} \cite{Dell1986, Mackay1988, Grainger1996}, которые заполнили `блоки' одиночными искусственными нейронами, которые, как предполагалось, могут локально описывать содержание опыта \cite{Barlow1972}. Взаимно-однозначное соответствие между искусственными нейронами и функциональными модулями делает крайне элементарным переход от блочных моделей к нейросетевым, однако, предположение о том, что отдельные нейроны могу быть ответственны за целостные функции не выдерживает никакой критики \cite{Abeles1991, Quiroga2019}. Помимо этого, подобные модели не позволяют моделировать процессы обучения и не дают никакой динамики поведения моделируемой системы. 

\subsection{Аттракторные нейросетевые модели}

Благодаря нейроанатомическим исследованиям коры головного мозга человека, было обнаружено, что связи в ней характеризуются высокой плотностью внутренних контактов и множественными повторами межнейрональных связей. Данное наблюдение позволило исследователям предположить, что такая архитектура связана с ассоциативной памятью \cite{Braitenberg1978, Braitenberg2013}. Эта гипотеза послужила основой для создания семейства нейросетевых моделей называемых `аттракторными нейросетями'  \cite{Willshaw1969, Lundqvist2006, Lansner2009, Hopfield1986}. Аттракторные нейросетевые модели предполагают, что нейроны имеют контакты с большинством или со всеми нейронами сети, что разительно отличается от архитектуры, используемой в стандартных искусственных нейросетях, о которых речь пойдет далее. Для реализации процессов обучения в аттракторных сетях были сформулированы правила обучения, которые позволяют менять веса на контактах между нейронами ориентируясь на активность каждого отдельного нейрона в системе. Например, правило Хебба, которое позволяет реализовывать процесс обучения без учителя в биологических нейросетях \cite{Hebb1949}. Согласно ему, связь усиливается там, где пресинаптический нейрон активируется в некотором временном окне до активации постсинаптического нейрона. И, наоборот, ослабляется там, где активация постсинаптического нейрона во времени предшествует активации пресинаптического нейрона. Данное правило нашло экспериментальное подверждение в механизмах долговременной потенциации и долговременной депривации. Предполагается, что эти нейронные цепи и циклы, сформированные в процессе обучения по правилу Хебба, могут функционировать как распределенные сетевые представления перцептивных, когнитивных или «смешанных» контекстно-зависимых перцептивно-когнитивных состояний \cite{Rigotti2010, Huyck2013, Lindsay2017}. Следовательно, то, что корковые нейроны работают совместно в группах \cite{Abeles1991, Quiroga2019} и что функционально они распределены по тем же группам, может быть смоделировано аттракторными сетями. Это позволяет моделировать процессы памяти и обучения, привнося видимое преимущество в сравнении с локальными нейросетевыми моделями. 

В аттракторных сетях возможна полная активация нейронального ансамбля, за счет лишь частичной стимуляции, что аналогично гештальту: восприятие объекта возможно по частичной/зашумленной части. Подобная устойчивость достигается за счет сетевой архитектуры, которая стремится вернуться к заученному состоянию. Примером такого типа модели может служить \textit{сеть Хопфилда} \cite{Hopfield1984, Hopfield1986}. Сеть Хопфилда позволяет моделировать механизмы ассоциативной памяти. Экспериментальные данные показывают, что даже при половине вышедших из строя нейронов в сети вероятность правильного ответа на стимул стремится к $1$. Сеть Хопфилда может состоять из $N$ нейронов, каждый из которых имеет аксо-дендритную связь с любым другим нейроном сети. В каждый момент времени $t$ всякий нейрон может находиться в одном из двух состояний $S_i(t) \in \{-1,1\}$, ответственных за модельное `возбужденине' и `торможение'. Динамика во времени $i$-го нейрона описывается дискретной динамической системой

\begin{equation}
    S_i(t) := \text{sign}[\sum_{j=1}^N J_{i,j}S_i(t-1)],
\end{equation}

где $i,j \leq N$ и $J_{i,j}$ — матрица весовых коэффициентов, определяющих взаимодействие дендритов $i$-ого нейрона с аксонами $j$-ого нейрона. При этом в сети отсутствуют петлевые связи -- действие нейрона на себя $J_{i,i} = 0$. Обучение такой сети сводится к задаче минимизации некоторого функционала и подбору значений матрицы $J$. 

Аттракторные нейросетевые модели могут быть использованы для моделирования широкого спектра нейробиологических феноменов и когнитивных процессов: от узнавания объекта/слова до опеределения пути и принятия решений \cite{Willshaw1969, Palm2014, Hopfield1986, Hinton1991, Rigotti2010}. Некоторые модели даже пытаются использовать наборы взаимодействующих друг с другом аттракторных нейросетей для моделирования сложных когнитивных процессов \cite{Papadimitriou2020, Dominey1995, Bibbig1995, Knoblauch20021, Knoblauch2002}. Более того, аттракторные нейросети могут быть включены в архитектуру более сложных нейросетей, например, глубоких, которые будут обсуждаться далее. 

\subsection{Многослойные персептроны}

В отличие от аттракторных моделей, где сеть имела вид полного графа, в случае с \textit{многослойными персептронами} архитектура сводится к наличию контактов только между нейронами из соседних слоёв \cite{Minsky1988}. Такие сети состоят из  нейронов соединенных последовательно, слой за слоем. Всего слоёв выделяют три: входной, скрытый и выходной, что навеянно нейроанатомией сетчатки \cite{Hubel1995}.

Такие модели дают представление объектов `плотно' упакованным и не разреженным: они распределены по всем нейронам скрытого слоя, так, что \textit{вектор активации} по всему скрытому слою является нейросетевым коррелятом представленного объекта, слова, значения или мысли \cite{McClelland1985}. Плотное распределение сигнала по сети разительно отличается от того, что мы видим в аттракторных или спайковых нейросетях. Обучение же реализуется методом градиентного спуска, когда после каждого прохождения сигнала через сеть результат сверяется с эталоном, считается мера ошибки и относительно нее корректируются все веса в сети, которые характеризуют собой силу синаптических контактов внутри модели \cite{Rumelhart1986}. 

По мере обнаружения ограничений данной модели для реализации механизмов памяти, было предложено добавлять в архитектуру дополнительные слои \cite{Elman1996}, что создает реверберации в активности сети и расширяет потенциальную применимость в зачах моделирования когнитивных процессов \cite{Rumelhart1986, Rogers2004}.

\subsection{Глубокие нейросетевые модели}

Дальнейшее развитие пошло по пути добавления большего числа слоев в многослойный персептрон, что получило развитие в теории глубоких нейросетей \cite{Richards2019, Hinton2007}. Нейробиологическая мотивация, стоящая за идеей все большего и большего увеличения числа слоёв в нейросети состоит в том, что похожую нейроанатомическую структуру имеет наша зрительная система (теперь уже не только сетчатка, но зрительная кора) \cite{Felleman1991}. 

Глубокие нейросети претерпели череду эволюций, приведших к появлению несовпадающих архитектур. Например, появились сверточные нейросети, которые включают топографические проекции между (не обязательно всеми) слоями сети, чтобы упрощать процесс обработки смежных входных данных: для различных нейронов выходного слоя используются одна и та же матрица весов, которую также называют \textit{ядром свёртки}, что отличается от наборов индивидуальных весов у каждого нейрона в стандартной модели глубокой нейросети. Подобные улучшения и модификации структуры многослойного персептрона позволили достичь человеческого уровня (качественного) решения таких задач, как классификация объектов \cite{Krizhevsky2012} или распознование речи \cite{Dahl2011}. 

Однако, при всём том успехе, которого достигли данные модели в решении конкретных задач, в них обнаружен целый ряд проблем. Первое, что стоит отметить, это направленность на решение одной узкой задачи и невозможность сочетать несколько разномодальных задач в одной сети. Глубокие нейросети обнаруживают в себе тенденцию к неуместным обобщениям, например, в задачах классификации, выдавая ответ на слишком сильно зашумленных и неидентифицируемых данных \cite{Szegedy2013, Nguyen2019}. Во-вторых, существует целая серия работ о неустойчивости подобных моделей к некоторым незначительным пертурбациям и шумам во входных данных, которые человек с легкостью бы отфильтровал \cite{Szegedy2013, Carlini2017}. В-третьих, такие сети не имеют динамики и активируются также плотно, как и обычные многослойные персептроны, что разительно отличается от принципов нейросетевой активности в мозге человека, которая имеет разреженную природу. Недавнее исследование показало, что введение скрытого слоя, подражающего структуре слоя V1 зрительной коры приматов в архитектуру глубокой нейросети, повысило устойчивость нейросети к шумам во входном сигнале \cite{Dapello2020}. 

Сейчас существует множество нейросетевых архитектур и почти все они основаны на изначальной идее персептрона. Варьируется активационные функции, функции ошибки, число слоев, но принципиальные проблемы этим не решить. Подобные нейросети как не имели динамики во времени, так и не получат её без кардинального изменения подхода к архитектуре модели \cite{Cazin2019}.

\begin{center}
\textit{\large Сведения об авторах}
\end{center}

\textbf{Онучин А.~А.}, студент 5-го курса факультета Психологии МГУ. 
\\
\textbf{Телефон:} 89851103522
\\
\textbf{Email:} arseniyonuchin04.09.97@gmail.com

\end{document}